\documentclass[
aip,
 amsmath,amssymb,
 reprint,%
]{revtex4-1}

\usepackage{graphicx}
\usepackage{dcolumn}
\usepackage{bm}
\usepackage[dvipsnames]{xcolor}

\usepackage[utf8]{inputenc}
\usepackage[T1]{fontenc}
\usepackage{mathptmx}
\usepackage{etoolbox}

\makeatletter
\def\@email#1#2{%
 \endgroup
 \patchcmd{\titleblock@produce}
  {\frontmatter@RRAPformat}
  {\frontmatter@RRAPformat{\produce@RRAP{*#1\href{mailto:#2}{#2}}}\frontmatter@RRAPformat}
  {}{}
}%

\newcommand{\ffcol}{\color{black}}  
\newcommand{\egcol}{\color{black}} 
\newcommand{\mlcol}{\color{black}}

\makeatother
\begin{document}

\preprint{AIP/JVST-B}

\title[Molten salt ion source using glass capillaries as emitter]{Molten salt ion source using glass capillaries as emitter}


\author{M. Léger}
\email{massimo.leger@tescan.com}
\affiliation{Centre de Recherche sur les Ions, les Mat\'eriaux et la Photonique (CIMAP), Université Caen Normandie, Normandie Univ, ENSICAEN, UNICAEN, CEA, CNRS, CIMAP UMR6252, F-14000 Caen, France}%
\affiliation{Orsay Physics, TESCAN Group, F-13710 Fuveau, France}

\author{E. Giglio}
\affiliation{Centre de Recherche sur les Ions, les Mat\'eriaux et la Photonique (CIMAP), Normandie Univ, ENSICAEN, UNICAEN, CEA, CNRS, F-14000 Caen, France}%

\author{S. Guillous}
\affiliation{Centre de Recherche sur les Ions, les Mat\'eriaux et la Photonique (CIMAP), Normandie Univ, ENSICAEN, UNICAEN, CEA, CNRS, F-14000 Caen, France}%

\author{A. Houel}
\affiliation{Orsay Physics, TESCAN Group, F-13710 Fuveau, France}

\date{\today}


\begin{abstract}
{\egcol 
A proof of concept of an ion source using a molten salt mixture rich in Cs and O ions, able to emit an ion current both in positive and negative polarities, is presented. 
 Similar to liquid metal ion sources, the ions of the salt are emitted by field evaporation from a Taylor cone formed under the influence of an intense electric field. In contrast to an "open" source design in which the conducting liquid covers the surface of a metallic needle, the presented setup uses a conical glass capillary as an emitter tip. {\ffcol The latter is observed by a camera, which revealed the formation of a bubble in the tip containing the molten salt. A glow was also observed during beam emission revealing the anchoring point of the stressed meniscus.} A stable ion beam emission of about 50 min and an intensity of several $\mu$A was obtained. The behavior of the molten salt mixture in  vacuum conditions is discussed, as its properties are essential to the design of the presented ion source.}
\end{abstract}

\maketitle

\section{\label{sec:level1}Introduction}

In Secondary Ion Mass Spectrometry (SIMS), charged particles are ejected from a sample surface when bombarded by a primary beam. Those charged particles form a secondary ion beam that is analyzed with appropriate electrostatic lenses to derive the charge-to-mass ratio of the particles.  The molecular mass spectrum of the pulverized area on the sample allows then for highly sensitive detection and imaging.
%
As a primary beam source and depending on the analysis requirements, typically O$^{2+}$, O$^-$, Cs$^+$, Ga$^+$ ions or clusters such as Bi$_{3}^{+}$ are used. While gallium ion sources are vastly used in  high lateral resolution SIMS when there is a need for fine image resolution, their excellent optical characteristics come unfortunately at a cost of ionization yield, i.e. the amount of secondary ions generated per pulverized atom: typical orders of magnitude for applications in metallurgy are around $10^{-5}$ with a Ga$^+$ beam, while yields of $10^{-3}$ using O$^{2+}$ or even $10^{-2}$ using Cs$^+$ beams are commonly achieved \cite{Schuhmacher,MIGEON199551}.

 An ideal primary ion source for Secondary Ion Mass Spectrometry (SIMS), would combine the brightness and low energy spread of a Liquid Metal Ion Source (LMIS), \textit{i.e.}, the emitted ion current per unit solid angle, with the availability of reactive elements as primary ions like Cs$^+$ beam or non-metallic ones like Oxygen or reactive clusters. 
Additionally, having the versatility of a duoplasmatron to produce both positive and negative ions would also be appreciated. From there, adapting LMIS source designs to work with other liquids containing different ion species naturally comes to mind.

{\egcol 

In order to focus ion beams into a spot size less than 10 nm, the beam needs to be emitted from a very small area into a limited solid angle. To meet this demand, several LMIS-type sources using a  needle or a capillary with micrometer tips have been proposed. In both cases, the conducting liquid at the tip of the emitter forms a Taylor cone under a strong electric field. The high field at the cone's apex extracts the ions and accelerates them, producing a bright beam. To sustain a steady beam, the liquid is fed passively to the tip, relying on capillary action along externally wetted needles, inside capillaries, or through bulk porous media. The needle-type emitter is most commonly used in Focused Ion Beam (FIB) instruments.

Few Cs-based LMIS designs have been reported \cite{KROHN_1975,PREWETT_1984,UMEMURA_1992,KAORU_1994}, arguably due to the chemical activity, low surface tension, and high vapor pressure of cesium, as opposed to the  successful Ga-LMIS. In fact, liquid caesium has to be supplied to the emitter / reserve by breaking a cesium vial under ultra-high vacuum (UHV) to prevent oxidation of Cs \cite{PREWETT_1984,KAORU_1994}, so that Cs-LMIS designs are difficult to handle or suffer a short lifetime and therefore are mainly suited for academic purposes.

Instead of pure liquid metals, Liquid Alloy Ion Sources (LAIS) use liquid alloys based, for example,  on gold-silicon eutectics to produce ion beams of different species, see L. Bischoff \textit{et al.} for an extensive review \cite{Bischoff_2016}. While LAIS sources are able to emit numerous ions like Li, Be, Ge\ldots, Cs-emitting LAIS sources have, to our knowledge, not been reported, probably because they are as challenging to operate as the pure Cs-LMIS. 

Liquid metals are not the only substances from which ion beams can be obtained. 
Ionic liquids (IL) have  been investigated as alternatives to LMIS  due to their low vapor pressure, chemical stability, low operating temperature and tunability via selection of cation/anion pairs. 
In the pioneering work by Romero-Sanz \textit{et al.} \cite{Romero_2003}, it was shown that positively charged solvated ions could be extracted from the ionic liquid [EMI]$^+$ [BF$_4$]$^-$ in a pure ion evaporation regime using silica capillaries as emitters. 
 Martino \textit{et al.} demonstrated using the capillary rise method that both hydrophilic and hydrophobic ionic liquids can wet borosilicate glass in dry conditions \cite{Martino_2006}. Hence, ILs should also be suitable for use with borosilicate glass capillaries as emitters. 

}
{\ffcol
Ionic Liquid Ion Sources (ILIS) using a tungsten emitter tip wetted by room temperature ionic liquids were investigated among others by Lozano and Martinez-Sanchez 
\cite{LOZANO2005, LOZANO2004149,Perez-Martinez_2010,PEREZMARTINEZ20112088,PEREZMARTINEZ2012,PEREZMARTINEZ201413}. ILIS share qualities with LMIS sources necessary for focused ion beams (FIB) operation, such as pure ion emission and low energy spreads. These sources work at lower currents, room temperature, and are able to produce both positive and negative ions from many ionic liquids.

However, organic ILs present specific issues in SIMS: large organic fragments can lead to complex, overlapping peaks in mass spectra, making data interpretation harder. Larger organic ions may not efficiently sputter sample material or may fragment unpredictably. 
Also, ILIS operation in a single polarity is limited by the  formation of a charge double layer at the surface of the tungsten needle that is wetted by the ionic liquid, leading to emission degradation within a few minutes \cite{LOZANO2004149}. In the particular case of a porous tungsten emitter needle, continuous DC operation leads even to the degradation of the emitter tip \cite{Brikner_2012}. Now, the electrochemical degradation of a porous needle was mitigated by using a distal electrode that polarizes the liquid, while the emitter is electrically floating \cite{Brikner_2012}. The formation of a charge double layer could be avoided by alternating the polarity of the source at a rate of 1Hz \cite{LOZANO2004149}. This way, the beam could be stabilized for over 200 h, but such a mode is  generally not compatible with FIB operation. More recently, Fujiwara reported that a negative cluster ion beam could be produced continuously using hydrophilic ionic liquid, propylammonium nitrate, and that electrochemical reduction of the ionic liquids forms volatile products,  enabling continuous negative ion beam production \cite{Fujiwara_2020}. Irradiating the surface of an electrically insulated material, Fujiwara \textit{et al.} showed that the charging voltage is as low as 1 V, so that secondary ion mass spectrometry of an insulated organic sample is possible using the negative ion beam without charge compensation \cite{Fujiwara_2021}. While ILIS may potentially enhance the capabilities in FIB applications, the integration of ILIS into Fib-SIMS systems is still under development.

}

Some cesium salts with bulky organic anions, such as [Tf$_2$N]$^-$ are known to form low melting (125$^\circ$C) ionic salts \cite{Stritzinger_2018} and can even be mixed with other alkali organic salts sharing the same anion to lower the melting temperature of the eutectic mixture \cite{Hagiwara_2008}. They can also be mixed with organic ionic liquids containing, for example, organic cations like imidazolium, to form room temperature ionic liquids \cite{MONTI_2014}. While there seems to be no published work about a ILIS  based on organic Cs salts, this may be a promising way to produce Cs$^+$ ions with an ILIS.  

Alternatively, a bright Cs$^+$ ion source, which is not based on the field evaporation mechanism from an emitter tip, has been developed in the last years. Low-temperature ion sources (LoTIS) use magneto-optical trapping with laser cooling to create an intense atomic beam, which is then photoionised to produce a beam with a high brightness and a low energy spread \cite{Steele_2017,Li_2024,VITEAU201670}. Such a Cs$^+$ has been recently integrated with SIMS instruments \cite{Wirtz_2024}. 
However, compared to ILIS-type sources (including the one presented here), LoTIS systems generally produce much lower ion currents compared to ILIS. This limitation makes them less suitable for applications that demand high throughput for fast imaging time or for high-dose ion implantation. Also, implementing magneto-optical trapping requires complex laser systems, precise frequency stabilization, and carefully aligned optics. These components add to the system’s complexity, cost, and maintenance compared to the relatively straightforward setup of LAIS or ILIS. Finally, LoTIS works exclusively in positive emission mode.





Compared to organic ionic liquids, molten salts have a significantly higher electrical conductivity and surface tension, albeit at the cost of a higher melting temperature. 
Based on the finding of Garoz \textit{et al.} that the purely ionic regime in ILIS is achieved merely for liquids with high surface tension and high electrical conductivity \cite{Garoz_2007}, molten salts should be promising alternatives, if purely ionic regimes are required. Using several eutectic salt mixtures in their ILIS source, mostly alkali-chlorides but also a few alkali-nitrates and akali-bromides, R. Sailer and P. Sudraud \cite{Sailer96} achieved stable beam emissions with a lifetime of typically 20 hours, the lifetime being limited mostly by the evaporation rates of the ionic compounds. The emitter tip was a 500 µm alumina fiber machined into a conical shape with typical curvature radii below 50µm. The source could be operated for several dozens of hours in a single polarity without noticing any change in the emitter or emission characteristics. Time of flight measurements showed that the beam contained a variety of singly charged ions or small clusters. %

Tajmar \textit{et al.}, manufactured and tested an indium liquid metal ion source using a tapered silicate glass capillary with a sub-micrometer tip as an emitter tip \cite{Tajmar}. They demonstrated that a pure glass-based LMIS may form the basis for future LMIS-chips with very high current densities. From the current voltage curve, they deduced a source impedance of 19 M$\Omega$ and a tip radius of about 0.5 µm, which was the outer diameter of the glass tip. However, because the liquid metal does not wet the silicate glass, they had to apply an overpressure of 20 bar to the liquid to push it into the capillary tip. When an electric field was applied at the tip, they succeeded in emitting a stable beam of 20 µA from the liquid meniscus anchored at the tip of the capillary for about 800 s before the tip of the capillary was destroyed by the overpressure of the liquid and the Coulomb pressure applied by the electric field at the tip of the capillary.

Following the idea of Tajmar, we present a proof of concept of a tapered glass capillary ion source, intended for SIMS applications using a molten salt mixture rich in Cs$^+$ ions and negative oxygen complexes like NO$_3^-$.
In contrast to an "open" source design in which the conducting liquid covers the surface of a tungsten needle, the presented setup uses a closed design which should avoid the problem of limited flow supply to the tip and also limit the surface of molten salt that is exposed to the vacuum.
We also discuss in this work the fusion temperature, evaporation, wetting behavior and hygroscopicity of the molten salt mixture in vacuum conditions and how those properties influenced the design of the ion source body.

\section{\label{sec:elementsIonSource}Elements of the ion source}

\subsection{\label{subsec:moltenSalt}The molten salt}

{\mlcol

A salt containing Cs and O ions with a fusion temperature preferably well below 200$^\circ$C would be ideal, as some parts of the source are not designed to handle higher working temperatures.
Additionally, the salt must also have a sufficiently low vapor pressure when heated above the melting temperature, as the main limitation encountered by Sailer et al. was the ionic compound depletion due to high loss rates, ranging typically from 1 to 100 mg per hour for their tested salts \cite{Sailer96}. Although having the salt enclosed in a capillary instead of   fully open emitter would limit most of the evaporation losses, some caution might still be necessary when choosing the salt compound, as a high evaporation rate could mean an non-negligible gas phase near the emissive meniscus, possibly leading to important beam-gas interaction and thus emission instabilities or a degraded energy spread.


 }

An eutectic mixture of 53\% mol. CsNO$_3$  and 47\% mol. LiNO$_3$ respectively was prepared. 
The mixture is characterized by a melting temperature of ${173^\circ}$C at atmospheric pressure and a density of about $\rho= 2.5$ g/cm$^3$ \cite{Janz, Janz_correl, KOUASSI2021158131}. As a matter of fact,  adding LiNO$_3$ to CsNO$_3$ allows lowering the melting temperature of the latter ($414^\circ$C) significantly. It also raises the electrical conductivity of the liquid compound. In \cite{Janz}, the presented data shows that our mixture has an electrical conductivity of 75 S/m, around 50\% higher than pure CsNO$_3$ at any tested temperature.  In return, the eutectic mixture is more hygroscopic than pure CsNO$_3$,  which is a serious constraint that we will discuss later.

\paragraph{Evaporation rate under Vacuum :}

To check whether the evaporation rate of the mixture meets our requirements, a glass cover with a few mm was placed above the crucible containing the molten salt and let it be exposed to the vapors of the molten salt at $150^\circ$ C for more than 500 hours at a pressure of $3 \times 10^{-7}$ mbar. When we checked the  surface of the glass cover, we found no visible layer of salt on the glass.
We also noted that  the salt mixture  melts in vacuum at a temperature of 134$^\circ$ C, which is  significantly lower than the one measured at atmospheric pressure, namely 170$^\circ$ C.  While not exactly a surprise as most substances tend to have lower melting points with decreased pressure, this was a welcomed discovery as at lower temperatures the molten salt is less corrosive which ensures a better longevity of the source. We also found borosilicate glass to be suited for containing the salt mixture for extended periods of time, as no visible alteration either of the salt or of the capillary surface was observed.

\paragraph{Hygroscopicity of the salt mixture :}

Our eutectic mixture is quite hygroscopic, as it absorbs at a high rate moisture from the atmosphere. It becomes sticky after being exposed to the atmosphere for only 15 minutes, and turns into a saturated water-based solution of itself in less than 12 hours. Although the absorbed water may not pose a problem when  the salt is melted at atmospheric pressure, we found that the moisture poses a problem when melting the salt at vacuum pressure. In fact, when the solid mixture is heated beyond its melting temperature in primary vacuum pressure, the liquid salt begins to bubble heavily, releasing the absorbed water.
The bubbling gives rise to projections of salt that soil the chamber if not contained. We observed that with decreasing pressure, the bubbling strengthens in that larger bubbles are formed, although at a lower rate.
A complete desiccation of the molten salt at pressures below 10$^{-6}$ mbar may take up to 24 hours, but was found to be crucial if one intends to use molten salt inside micro-capillaries,
which is the case here.


%



\paragraph{Wetting behavior: }
The wettability of the  borosilicate glass capillary by the molten salt plays a crucial role in ensuring continuous feeding of the tip in order to sustain a constant beam emission from the Taylor cone. We placed a capillary glass tube (the same that  are used in the puller) and a crucible with a salt mixture in the oven at ${200^\circ}$C. Once the salt was molten, the hot capillary  was dipped vertically into the molten salt. The salt rose in the tube by $h \simeq 2.5$ mm above the free surface. Using Jurin's law $\cos \theta  = \frac{ \rho \, g \,d}{4 \gamma } h$ with the recommended value of $\gamma=0.11$ N/m \cite{Janz_correl} for the  surface tension of the salt at ${200^\circ}$C, we get an  estimated contact angle of $82\deg$, indicating that the molten salt wets only moderately borosilicate glass at atmospheric pressure. However, in the vacuum chamber, the molten salt will  be at pressures below $10^{-6}$ mbar. We thus checked  whether the molten salt also wets borosilicate in vacuum. The salt was first desiccated with the procedure discussed before. Then, still under vacuum condition, a glass tube was dipped into the molten salt. The height of the salt column was depressed in the capillary, indicating that desiccated salt does not wet borosilicate glass. It seems thus that the adsorbed water is responsible for the moderate wetting behavior of molten salt at atmospheric pressure. We checked other salts, containing potassium or  sodium atoms or salts containing chlorides instead of nitrates, but once the salts were desiccated, the same non-wetting behaviors of glass were always found. In the following section, we will see how this non-wetting behavior of glass was mitigated in our setup.


\subsection{Emitter tip}

\begin{figure}
\includegraphics[width=1.0\linewidth]{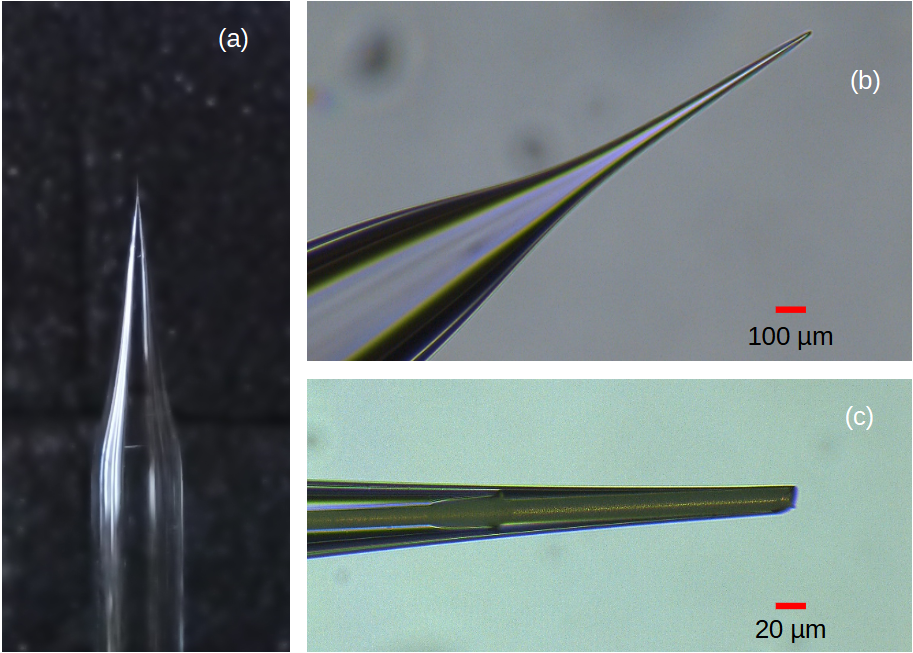}
\caption{\label{fig:shape}(a)  Shape of  a glass capillary after being pulled. The tapered part is about 4 mm long  and the outer diameter of the cylindrical part is 2 mm. (b) Zoom on the capillary tip.  (c)  Capillary tip, cut so as to insert a 20 µm alumina fiber.}
\end{figure}

The tapered glass capillaries we used in the ion source were obtained by pulling a locally heated  glass tube using a home-made capillary puller.  
An example of the resulting funnel-shaped tip is shown in Fig.\ref{fig:shape} (a). The tapered part is about 3-4 mm long, with an outlet diameter typically below 3 $\mu$m.  The zoom in Fig.\ref{fig:shape} (b) details the shape of the capillary tip.
We tested two different configurations for the capillary.


\subsubsection{Configuration one: Glass tip}

In the first configuration, the beam is supposed to be emitted from the meniscus of a Taylor cone anchored on the tip of a tapered glass capillary  (see inset (b) of Fig.\ref{fig:shape}) that contains the molten salt, similar to the concept tested by Tajmar \cite{Tajmar}. The original idea here was that the molten salt advances to the tip either by capillary forces or, if the salt does not wet sufficiently the glass, by applying a slight overpressure to the salt reservoir to overcome the first wetting. Once the liquid is present at the tip, it forms a pendant microsized drop, which, under the influence of an electric field, forms  a Taylor cone. Unfortunately, this configuration did not work properly because, as indicated before, the desiccated molten salt mixture does not wet borosilicate glass in vacuum and the salt did not fill the tip spontaneously. A force had to be applied to push the liquid all the way down to its tip, which was done by applying an overpressure in the molten salt tank, using typically an inert gas as N$_2$ or Ar, to push the liquid through the capillary towards its tip. 
The applied pressure must compensate for the negative capillary pressure, $ \frac{ 2 \gamma cos \theta }{r_{c}} $,
which scales with the inverse of the radius $r_{c}$ of the capillary tip. 
We typically needed to apply an overpressure of about 300 mbar to force the liquid up to a tip diameter of less than 3 µm.

\begin{figure}
\includegraphics[width=1.0\linewidth]{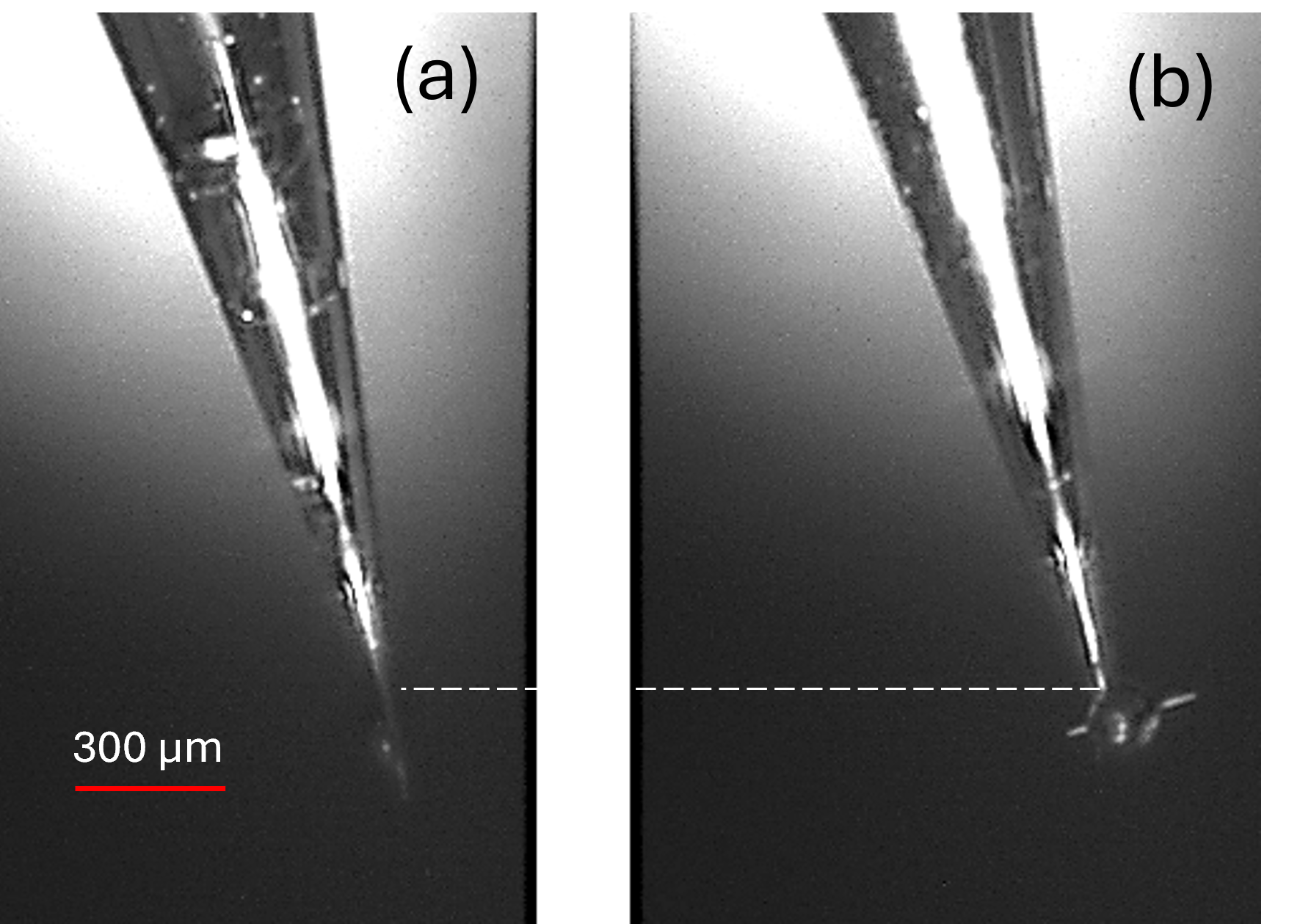}
\caption{\label{fig:tip_broken} (a) An overpressure of about 300 mbar pushes the molten salt into the tip. (b)  The broken tip is attached with a pending droplet to the capillary.}
\end{figure}

Fig.~\ref{fig:tip_broken}%
%
We found this filling method to be problematic, since the thin walls of the capillary are extremely fragile, and forcing the salt out of a micrometric tip often resulted in a broken tip, as was experienced by  Tajmar \cite{Tajmar} and as can be observed in Fig.~\ref{fig:tip_broken}. In the latter, the broken tip, attached by a pendent droplet to the capillary, is still visible.

\subsubsection{Configuration two: Glass tip with fiber}
\begin{figure}
\includegraphics[width=0.92\linewidth]{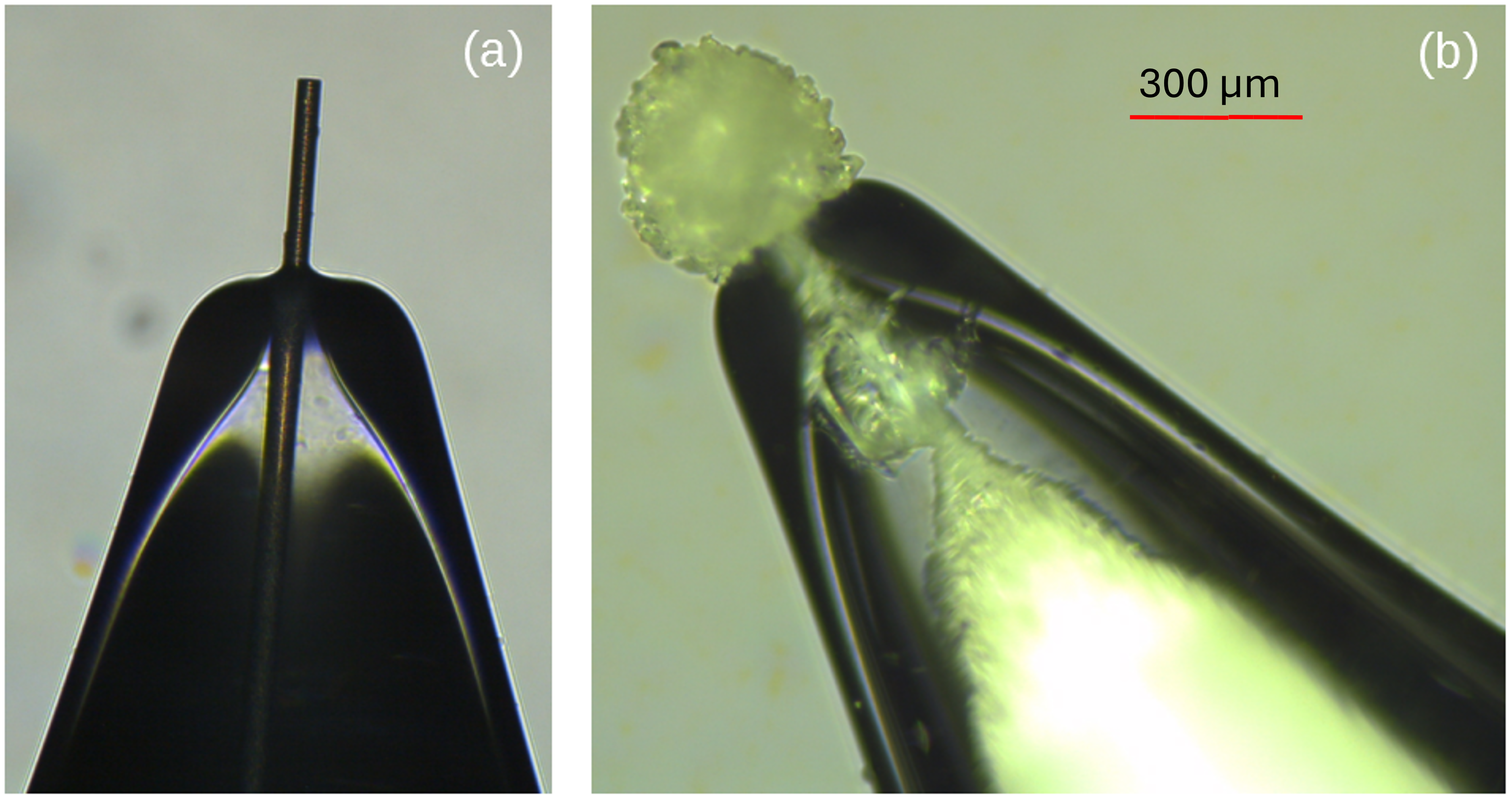}
\caption{\label{fig:wetting_salt_capillary} Wetting behavior observed with ethanol and molten salt. (a) Capillary with injected ethanol. Fiber's diameter is 20 µm. (b) Same capillary after injecting molten salt}
\end{figure}

The second configuration is motivated by the fact that our molten salt wets alumina in dry air as illustrated by the test described in the following lines. A capillary tip is cut so as to insert a 20 µm alumina fiber. We checked whether the capillary tip was not blocked by the inserted fiber by dipping the tip  into ethanol and observing the ethanol flowing inside the tip. (Fig. \ref{fig:wetting_salt_capillary}-(a)).
Then, salt is molten in a borosilicate beaker at 250°C in an oven and is injected in the pre-heated capillary still in the oven to avoid any thermal shock. The pressure is applied until some drops are expelled to ensure a homogenous and complete filling. The capillary is then brought back under the microscope to observe the salt.
Unlike ethanol that effectively wets both borosilicate and alumina, the molten salt clearly wets the alumina fiber preferentially over borosilicate glass (Fig. \ref{fig:wetting_salt_capillary}-(b)). 
The wetting of alumina by molten salts in dry air was also observed by Kirchebner \textit{et  al.} in the case of alkali-chloride molten salts \cite{Kirchebner_2021}.

So, in the second configuration, an alumina fiber is inserted into the glass capillary until the tip of the fiber is aligned with the tip of the capillary, without overshooting, see inset (c) of Fig.~\ref{fig:shape}. The fiber is then fixed  at the base of the capillary. In this configuration, the opening radius of the capillary tip is about 22 $\mu$m.  Where the diameter of the capillary approaches the diameter of the fiber, the salt is expected  be drawn along the fiber, potentially overcoming the lack of wettability of the capillary walls, until it reaches the apex of the tip, where it forms a stable droplet, see Fig. \ref{fig:capillary_experiment_run}(a).  
On the outside of the capillary, the non-wetting property of the borosilicate glass prevents the pending drop from spreading over the outer surface.   
%
The second configuration was used in the present work.
%

\subsection{The  ion source}

\begin{figure}
\includegraphics[width=0.92\linewidth]{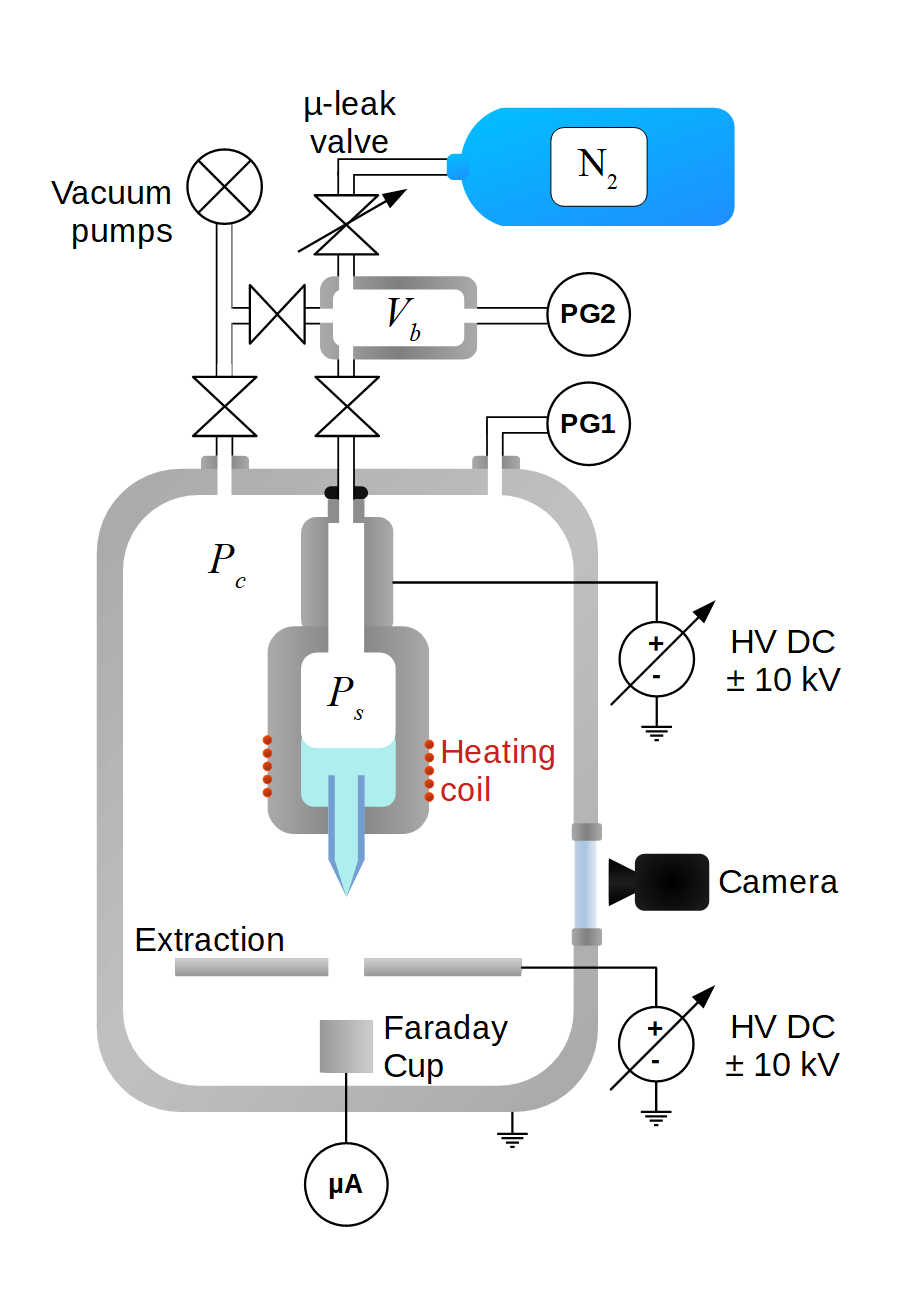}
\caption{\label{fig:source} Scheme of the ion source setup}
\end{figure}

The ion source body is a stainless steel salt container, with an embedded heating coil to melt the ionic compound inside. An outlet allows mounting the capillary in a liquid-tight junction.
The ion source is placed in a vacuum chamber of 1e-6 mbar during operation. Its mounting point in the chamber is an electrically insulating, airtight junction, so as to ensure two separate pressures in the chamber $P_{c}$ and inside the source $P_{s}$, see Fig.~\ref{fig:source}
The chamber pressure $P_{c}$ is monitored with a Pfeiffer PKR 251 vacuum gauge (PG1 in Fig. \ref{fig:source}). To control the pressure $P_{s}$ applied inside the  source body, a small vacuum chamber is used as a buffer volume $V_{b}$, where $N_{2}$ can be precisely injected using an adjustable µ-leak valve. Once the desired $N_{2}$ pressure $P_{s}$ is achieved, as read on a manometer (PG2 in Fig. \ref{fig:source}), injection is stopped, and the source valve is opened. The source's internal volume ($\sim$ 1 cm$^{3}$) being small compared to $V_{b}$ (a few liters), $P_{s}$ becomes indeed the pressure inside the source with negligible error. {\egcol While the setup allows adjusting the overpressure in the salt container, the overpressure is not finely controlled and thus not actively stabilized. }
Facing the capillary, an extraction electrode is placed at a distance of 2.5 mm from the emitting tip.
The source body and the extraction electrode are polarized using two high-voltage outputs of a FIB controller power supply unit (manufactured by Orsay Physics). A Faraday Cup (FC) allows collecting the central part of the emitted beam. 
%
%
Finally, a camera setup is mounted on a viewport ( Monochrome camera Flir \textsuperscript{\tiny\textregistered} BFS-PGE-63S4M-C \textsuperscript{\tiny\textregistered}, coupled with the following optical elements: Thorlabs \textsuperscript{\tiny\textregistered} MVL12X12Z, MVL12X20L, MVL20A  and MVLCMC ) to record the capillary, allowing to observe fluidic behavior inside the capillary and at its tip.  

\section{Results and discussion}
\label{sec:results}

\begin{figure}
\includegraphics[width=0.92\linewidth]{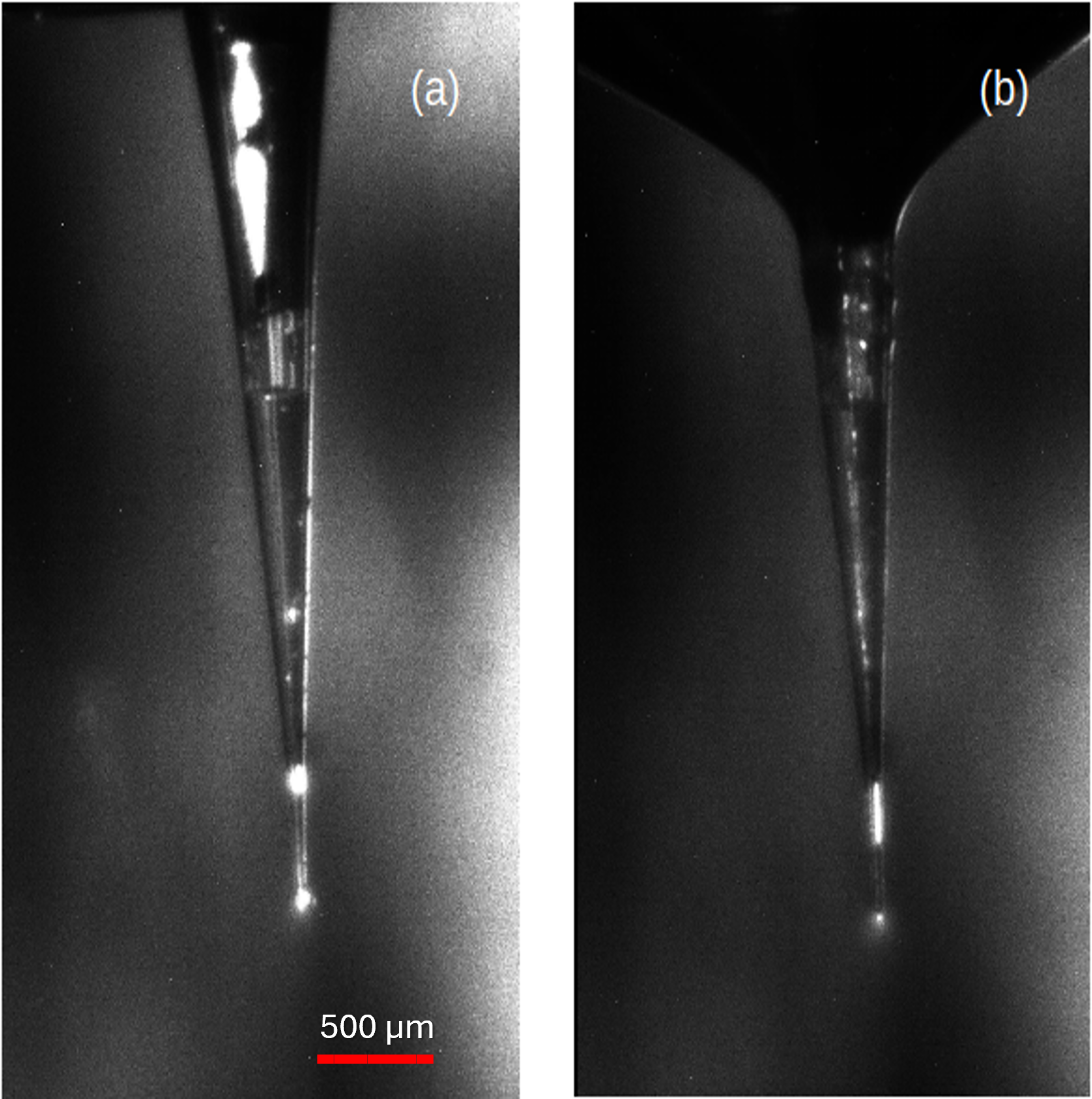}
\caption{\label{fig:capillary_experiment_run} (a) Capillary tip with molten salt and  alumina fiber (not visible) before applying the extraction potential. A pendant droplet of molten salt is visible at the apex of the tip.  (b)  During ion emission, a glow at the apex of the tip is visible.}
\end{figure}

 Although precautions to minimize the exposure of the salt to the atmosphere were taken, for example, by storing the solid salt between steps of manipulation in primary vacuum,  we did not succeed in avoiding that some moisture was absorbed by the salt during some steps that took place at the atmosphere, especially the last one where we mounted the source body on the vacuum chamber.  As a result, the salt contained in the capillary absorbed some moisture that was released in the form of submillimeter-sized bubbles once the salt was melted at vacuum pressures in the test chamber.  The submilimeter-sized bubble that separates the salt in two parts can be observed in 
 Fig.~\ref{fig:capillary_experiment_run}.
We expect that the electrical contact between the two separated parts is ensured by the molten salt wetting the surface of the alumina fiber. This assumption is further discussed in section below. 

  Before an electric field was applied, a pendant droplet of molten salt is visible at the apex of the tip, as was monitored by the camera and shown in the inset a of Fig. \ref{fig:capillary_experiment_run}. The droplet, which would hinder the source from working properly, is removed by applying an electric field that pulls the droplet until it detaches from the tip. 
  {\egcol The remaining liquid at the apex forms a meniscus  under the influence of the electric field, in equilibrium with the capillary pressure. Unfortunately, the camera lacks the necessary resolution to observe directly the shape of the meniscus at the apex. Coffman and subsequently Gallud \textit{et al.} have simulated numerically the shape of the  meniscus of an electrically stressed ionic liquid, anchored to the tip of a needle or of a capillary \cite{Coffman,Gallud_Lozano_2022,Gallud_phd_2023}. The exact shape of this meniscus is  dependent on conditions such as the space charge of the plume, the geometry of the emitter, or the liquid flow rate to the emitter tip. For a liquid anchored on the tip of a capillary and for an electric field just below the emission onset  and zero liquid pressure, it was found that the shape is slightly more elongated than the perfect Taylor cone solution, except a very small region at the apex with strong positive curvature, see figure 6.4 in \cite{Gallud_phd_2023}. 
  }
  
\begin{figure}
\includegraphics[width=\linewidth]{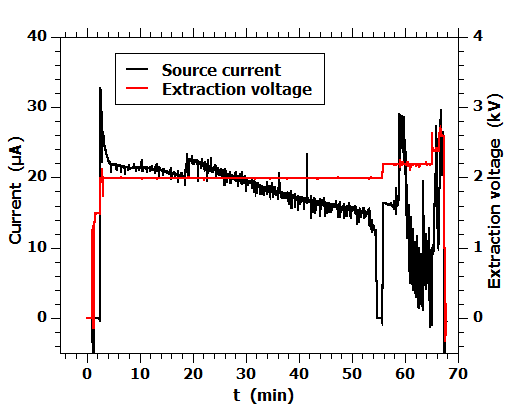}
 \caption{\label{fig:emission_50min} Emitted beam current (black line) and extraction potential (red line) as a function of time.}
\end{figure}

  Once the critical field is reached, ions are field-evaporated from the meniscus of the liquid anchored on the tip. At that moment, a glow at the apex is visible and a beam current is measured.  With increasing current, the luminescence intensity increases rapidly.
  {\egcol
  Uchizono \textit{et al.} investigated the emission spectra of glows produced by ionic liquid ion sources.  Their findings show that the glows at the emitter tip are not directly caused by the electrospray ionization process, but rather by electron excitation resulting from secondary species emission (SSE) impacting the surface of the emitter. The SSE, for their part, are generated by high velocity primary species that impact, for example, the surface of the extractor electrode. Thus, a significant glow should not occur in electrospray devices in the absence of secondary species \cite{Uchizono_Glow}. Other indirect mechanisms may also contribute to glow, including back streaming interactions between primary and secondary species, as well as field ionization and excitation of the residual gas molecules by the high electric field ($\sim 10^7$ V/m) near the emission tip. {\ffcol 
  In the case of LMIS, spectroscopic studies of the EHD emitters of gallium ions showed that the glow is mainly due to a cloud of excited gallium atoms \cite{Venkatesan_1981}. For Mazarov \textit{et al.}, one of the possible explanations for the appearance of a gas cloud,  is that the field evaporation of an ion is a nonequilibrium process, where it is possible to transfer energy to the nearest surface atom, leading to its evaporation. Of course, such “concomitant” evaporation occurs with low probability, but it is possible that it is responsible for the appearance of a small stream of neutrals in the form of a vapor with a high density \cite{Mazarov_2020}.}
  
  In the setup used here, the grounded extractor electrode with an aperture of 5 mm was placed 2.5 mm from the emitter tip. During beam emission, the most divergent part of the beam certainly hit the surface of the electrode, as a reddish glow ring surrounding the aperture could be observed through the window. It is thus reasonable to assume that the glow on the emitter tip is due to SSE, as proposed by \cite{Uchizono_Glow}.

  ILIS sources can operate in various emission regimes, depending mainly on the fluid flow rate. In the cone-jet mode, 
  the flow rate, must be sufficient to compensate for the emitted jet. Reducing the fluid flow rate at constant conductivity decreases the jet size, making the electric field strong enough for ion evaporation from the charged interface. This creates a mixed ion-droplet beam. Lowering the fluid flow rate further results in pure ion emission, characterized by the absence of droplet current \cite{Gallud_Lozano_2022}. While the observed glow does not allow distinguishing  between the three modes, we expect the source to operate in pure ion  mode because of limited flow rate as the molten salt does not wet the glass. Still, in the absence of time of flight measurements,  the composition of the beam remains speculative. 

 During the total 119 minutes of operation of the source, several emission tests were conducted, to observe the emitted current variations with the extraction voltage, and also to determine the drift in current when maintaining a constant extraction voltage. The longest emission had an intensity of about 21 µA and was roughly stable for about 50 minutes, albeit characterized by a steadily slow decrease over time, going from 21 to 15 µA, see  Fig.~\ref{fig:emission_50min}. }
%
 %
{\egcol
This tends to confirm the observations, reported by Sailer \textit{et al.}, that ILIS sources using molten salts with a dielectric emitter tip do not have to alternate the polarity to avoid the degradation of the beam \cite{Sailer96}. The electrochemical double layer at the electrode-liquid interface, that degrades the tip of the conducting emitter needle in DC mode, is certainly avoided here by the fact that the beam is emitted from the tip of an insulating capillary.

In Fig.~\ref{fig:emission_50min},  the beam intensity dropped abruptly after about 50 min of continuous emission. An increment of the extraction potential of 100 V could restore the beam for about a minute before it became unstable, fluctuating strongly, and eventually stopped again. Further attempts to increase the potential resulted in unstable onsets and discharges. 



Figure~\ref{fig:IV} displays the beam intensity as a function of the extraction potential (I-V) for both polarities.
From the curves, the  impedance of the source, which is equal to the inverse of the slope I-V, can be deduced. 
} {\mlcol
In positive mode, beam emission starts at around 1.4 kV with a source impedance of roughly 100 M$\Omega$, well above the 19 M$\Omega$ found by Tajmar on his indium-capillary source \cite{Tajmar}. The current then increases linearly up to 2.1 kV, before rising rapidly up to 20 µA, with an impedance of 20 M$\Omega$ around 2.5 kV. Rather than a change in emission regime, the shift in impedance is rather attributed here to a change of the anchoring point of the meniscus at the tip of the capillary. In Fig.~\ref{fig:meniscus_evol}, the four snapshots show that the location of the emitting meniscus has changed over time. Below 2.1 kV, the beam was emitted from a sharp and chipped edge of the capillary (the edge was not fire-polished), breaking the axial symmetry.
%
In negative mode, the onset was found around 1.6 kV indicating that the cohesion energy of NO$_3^-$ anions in the molten salt is slightly higher than for the cations. The measurements were limited to 2.3 kV so that the change in impedance was not observed. } {\egcol 


\begin{figure}
\includegraphics[width=1\linewidth]{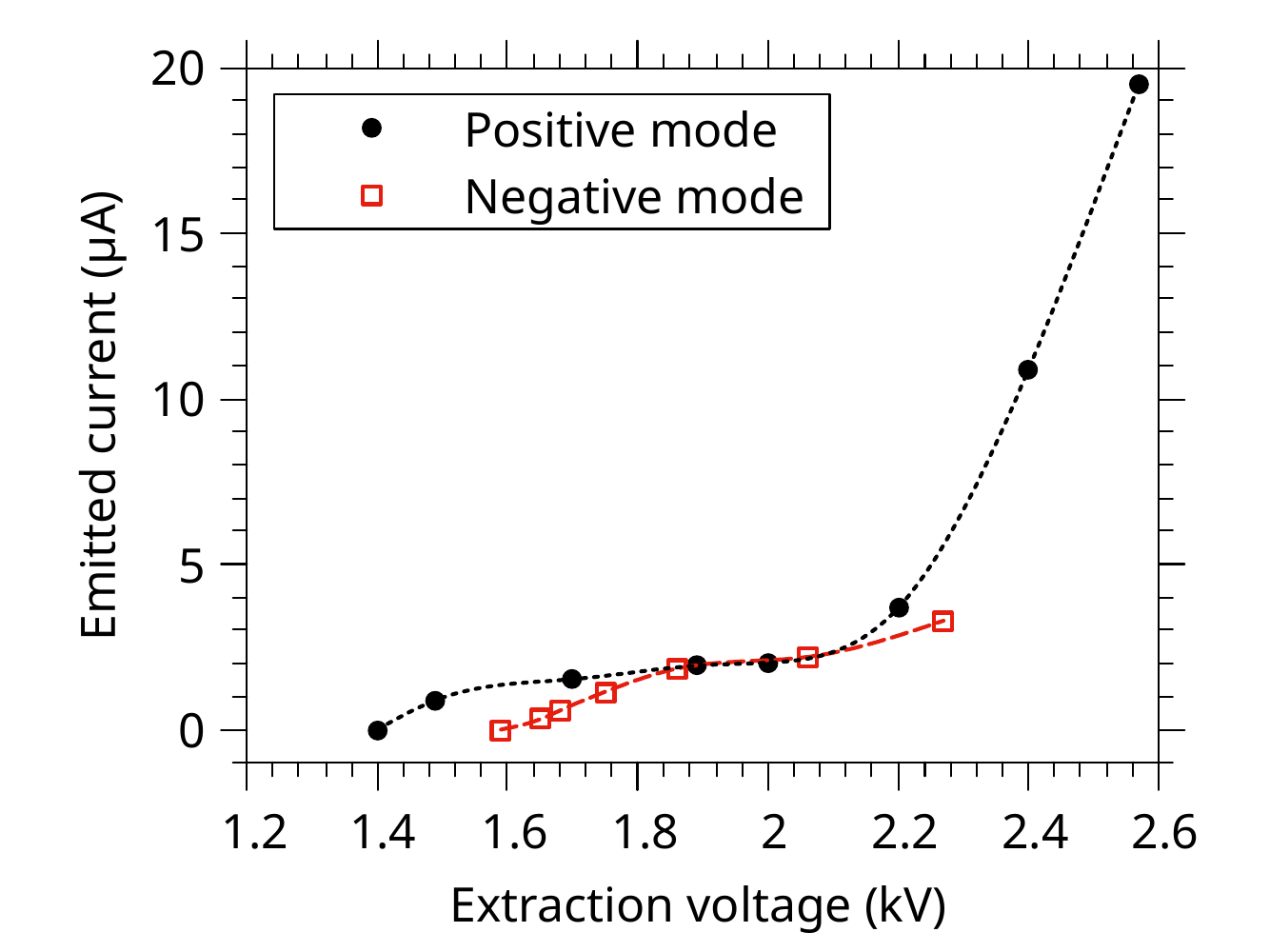}
\caption{\label{fig:IV} Emitted beam current in positive mode (black full circles) and negative mode (red squares) as a function of applied extraction voltage. The dashed curves  guide the eye.}
\end{figure}

}






In an attempt to explain the slight drift of the source current observed in Fig.~\ref{fig:emission_50min}, two scenarios are considered. 
Visibly, the lower salt column is separated from the upper one by a  bubble that formed after the salt was molten, presumably due to moisture absorbed during unavoidable manipulation steps in the atmosphere. If the lower salt column was electrically isolated from the potential of the source body, a continuously emitted constant current  of more than 20 µA during about 50 min could not be sustained, as the potential $U$ of the lower salt column would drop significantly. Indeed, modeling the lower salt column by a conducting cylinder of  length $L=1000$ µm and radius $R=50$ µm allows estimating  the electrical self-capacitance $C$ of the conducting salt column, $C\simeq 4 \pi  \epsilon_0  L/ (\ln (4L/R)-1) \sim 0.03$ pF \cite{Chalmers_1980}. The drop of the potential $\Delta U$ is given by the relation $\Delta U = \frac{1}{C} I \Delta t $, which means that for a current $I$ of about 20 µA, the potential of the lower salt column would drop by 1000V in less than 1 µs. We must thus assume that the lower salt column was electrically connected to the upper one, presumably via a liquid salt layer along the alumina fiber, which was able to compensate for the loss of the electric charge by the beam emission. One may now argue that the conductance of the liquid salt layer on the fiber could not completely balance the emitted current so that the potential $U$ of the lower column would eventually decrease slightly, explaining the steady decrease of the emitted current. But then one would expect that a steady  emission could be reestablished after waiting for several seconds without beam emission for the charges to relax. {\mlcol A steady current could however not be restored after such waiting periods, rendering this hypothesis unlikely.}

Another explanation would be that the slow decrease of the emitted current was related to a decrease of the pressure in the salt reservoir of the source body. Indeed, since the desiccated salt does not wet borosilicate glass, a small over-pressure of about 30 mbar had to be applied to the reservoir until the liquid salt fills the tip of the capillary. But nothing in the setup enforces the over-pressure to be constant over time, so it is possible that the pressure in the reservoir decreased  with time and that finally the molten salt was pulled back from the tip, preventing a sustained Taylor cone. The latter scenario is compatible with the observation that no stable emission could be restored, even after increasing the extraction voltage. {\mlcol It would also reinforce the assumption that the meniscus anchoring point changed over time. 
\begin{figure}
\includegraphics[width=0.92\linewidth]{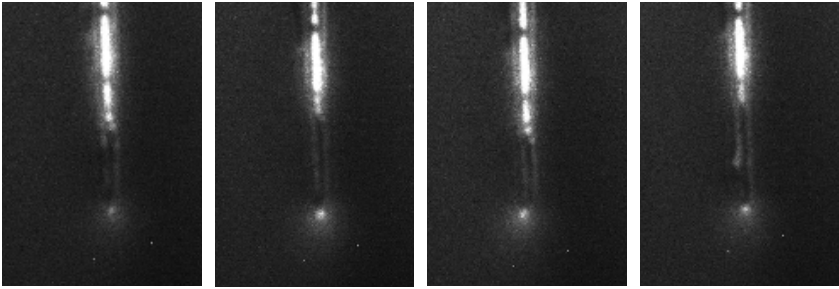}
\caption{\label{fig:meniscus_evol} Snapshots of the various glow location on the capillary tip.}
\end{figure}

}

\section{Conclusion}


We presented a proof of concept of an ion source that works both in positive DC mode and negative DC mode, emitting ionic current from a molten salt rich in caesium and oxygen. 
The beam was emitted from the micro-sized tip of the capillary with a current intensity of $\sim 20$ µA and was stable (except for the slight drift) for about 50 min before it stopped. The slight decrease of the beam current is discussed and possibly linked to the change of pressure in the salt reservoir. The camera that monitored the tip of the capillary is an imported addition to the setup, as it allowed us to see the bubbles that appear in the salt column  due to residual moisture in the salt, which eventually helped us to identify an unexpected problem  that needs to be tackled in future work. The camera also allowed us to monitor the glow at the apex of the tip during beam emission and to notice that the anchoring of the electrified liquid meniscus has changed over time.
The difficulties related to the hygroscopicity of the salt mixture and its low wetting behavior with respect to borosilicate glass were discussed, as well as a way to mitigate the latter by inserting an alumina fiber into the tip of the capillary to help the salts flow more easily to the tip. Additionally, the fiber ensures the electrical continuity between salt columns in the tip separated by microsized bubbles. In future works, it would be interesting to modify the surface energy of the glass by depositing, for example, an alumina layer at the inner surface of the capillary to ensure that the capillary tip is more readily wet by the salt mixture. That would  allow a sub-microsized tip to be used as an emitter. Upon a good wetting of the inner capillary surface by the salt, a stable ion beam can be expected. 
{ \ffcol Alternatively, because of limited wetting of the glass surface by the desiccated molten salt, the liquid is not passively fed to the tip of the emitter, and the overpressure in the salt reservoir must be actively controlled. In future work, the active control of the overpressure will be implemented in the setup to ensure a stable emission. Also, fire polishing the edge of pulled capillaries would certainly avoid emitting from chipped edges.  Eventually,  the source body with desiccated salt will be handled  and filled under glove box and inert gaz to minimize exposure to humidity. }

\begin{acknowledgments}
We thank Jean-Marc Ramillon (CIMAP) for manufacturing many mechanical parts of the setup. Acknowledgment for the financial support from the French National Agency (ANR) via the Common Laboratory LabCom CiCLOp (Grant No ANR-18-LCV3-0005-01)
\end{acknowledgments}


The data that support the findings of
this study are available from the
corresponding author upon reasonable
request.

\bibliography{CAPSEL_biblio}

\end{document}